
\input jnl
\centerline{\bf A Model for Solid $^3$He: II}
\vskip .5truein
\centerline{J. Dubois and P. Kumar}
\centerline{Department of Physics
and
National High Magnetic Field Laboratory}
\centerline{University of Florida}
\centerline{Gainesville, FL  32611}

\vskip .5truein
\centerline{\bf Abstract}
We propose a simple Ginzburg-Landau free energy to describe the magnetic phase
transition in solid $^3$He.  The free energy is analyzed with due consideration
of the hard first order transitions at low magnetic fields.  The resulting
phase diagram contains all of the important features of the experimentally
observed phase diagram.  The free energy also yields a critical field at which
the transition from the disordered state to the high field state changes from a
first order to a second order one.
\vfill\eject
\oneandahalfspace
\noindent{\sc I.  Introduction}

Solid $^3$He undergoes a phase transition$^{(1)}$ at $T_N=1mK$ into an
ordered state that consists of ferromagnetic (100) planes and two
successive planes with parallel spins followed by the next pair of
planes where the spins are antiparallel.  This state, as the magnetic
field is increased to approximately 0.5 T, is replaced by a normal
canted antiferromagnet.  The transition to the low field uudd
(up-up-down-down) phase is first order.  It is also first order to the
canted antiferromagnet phase for fields $B<0.6T$.  It becomes second
order$^{(2)}$ for $B>0.6T$.  The point $B=0.6T$ is similar to a
critical point.

The standard model$^{(3)}$ for solid $^3$He consists of pair, triple
and four spin exchanges analyzed within  a mean field theory.  The
model has been successful in deriving a large number of experimental
observations.  The multiple spin exchange model provides important
physical insights.  It shows that because of the competing interactions
(even spin permutations lead to antiferromagnetic and odd permutations
to ferromagnetic exchange interactions, see for example D. J. Thouless,
ref. (3)) both the Curie-Weiss constant and the bare Neel temperature
are small.  This feature for example
resolves a long standing disagreement between exchange constants
measured from $T_1$ measurements and from Curie-Weiss temperature.
While the former is an average of squares of the exchange constant, the
latter is a linear sum.  Exchange constants of different signs
contribute differently to the two observables.  It also shows that
solid $^3$He is a frustrated antiferromagnet.  There are some
difficulties.  For example the molar volume dependence of various
observables is almost identical, giving rise to suspicion$^{(4)}$ that
a much smaller number of energy scales are involved.  Quantum Monte
Carlo calculations$^{(5)}$ show a weak convergence in the magnitudes of
exchanges involving larger number of spins.  Finally, multiple spin
exchange models are sufficiently complex so that it often becomes
difficult to obtain a qualitative understanding of processes in solid
$^3$He, something that lately has become essential as more complex
phenomena are discovered.  For example, in thermal conductivity
measurements $^{(6)}$ there is evidence for magnetic defects and
we hope to be able to calculate the energies of these defects through
a Ginzburg-Landau model.

Our aim in this paper is to propose a Ginzburg-Landau type free energy.
The free energy is not---at least yet---derived from any microscopic
Hamiltonian.  Its terms are those allowed by the various ground states,
satisfying lattice symmetries, and interactions between them.  We study
thermodynamic consequences with the aim of fine tuning its terms to
produce a coherent, albeit phenomenological model for solid $^3$He.
Eventually, one task will be a microscopic calculation of the
parameters.

Our starting point is a proposal by Guyer and Kumar$^{(7)}$ (GK).
Noting that the spin susceptibility was close to Curie law just above
the transition at low fields, they argued that this implied a special
cancellation of the interaction energies.  They further argued that
the nature of the first order transition, hard in the sense that the
entropy discontinuity is a major fraction of $R ln 2$, further implied
that the free energy for solid $^3$He is given by
$$F= Jm_2^4- m_0B-T\Sigma (m)\eqno(1)$$
Here $B$ is the magnetic field, $T$ is the temperature and $\Sigma$,
the entropy given by
$$\Sigma(S) =\ell n 2 -{1\over 2} [(1+S)\ell n (1+S) +(1-S) \ell n
(1-S)]\eqno(2)$$

The magnetization $m(\twiddle r)$ is a vector of length $\mu_N S$ and
is given by
$$\twiddle m(\twiddle r) = \twiddle m_0 +\twiddle m_2 \cos (\twiddle
k_2\cdot \twiddle r + \pi/4)\eqno(3)$$

The vectors $\twiddle m_0\cdot \twiddle m_2=0$ and $|m|^2 = m_0^2 +
m_2^2$.  In our notation, when the vector sign is omitted, the quantity
refers to the amplitude of the vector.  Here $\twiddle k_2$ is half of
the reciprocal lattice vector in (100) direction and leads to the uudd
state.  The phase angle $\pi/4$ ensures that all spins are of the same
length.  The entropy $\Sigma(S)$ is the entropy of a spin $1/2$ system.
In general the entropy is a consequence of proper configuration counting
subject to the constraints on the length and the discrete nature of the
spins.  The expression in Eq. (2) reproduces the mean field equation
of state.  In the following we shall use the entropy as in Eq. (2).  Note
however that S is scaled so that its two possible values are $S^2=1$.
The effect of a quadratic term in $m_2$ has been discussed in
some detail in ref.~(8).

In a recent preprint,$^{(8)}$ we have argued that a Ginzburg-Landau
theory as an expansion in powers of the order parameter is inappropriate
(and inadequate) to describe a first order phase transition.  In a hard
first order phase transition, the order parameter, at the phase
transition jumps to a value which is easily outside the range of
validity of Ginzburg-Landau theory.  We cannot expand the entropy in
Eq.~(1) in powers of the order parameter and Eq.~(1), subject to the
constraint, $m^2=m_0^2+m_1^2+m_2^2$ must be solved, if necessary,
numerically as was already done in ref. (8).  While Eq.~(1) was
applicable only at low fields and near the phase transition, the
corresponding free energy for all fields is described in the next
section.

This paper is organized as follows:  in section II, we introduce the
free energy involving leading contributions of all of the possible
ground states.  Section III contains a solution of the phase diagram
for $B=T=0$ in the parameter space of the free energy.  Section IV
describes the results and we conclude with a summary and a discussion
of possible improvements.

\vskip .15truein
\noindent{\sc II.  The Model}

The minimal model should contain the three order parameters $m_0$,
$m_1$, and $m_2$ such that local magnetization is given by $\twiddle m
(\twiddle r) = \twiddle m_0 + \twiddle m_1 cos(\twiddle k_1.\twiddle r)
+\twiddle m_2 cos(\twiddle k_2. \twiddle r)$.  Here $k_1$ and $k_2$ are
respectively the wavevectors for the simple antiferromagnetic and the
uudd states.  The transition in $m_2$ is hard first order and we
consider only terms quartic in $m_2$.  The order parameter $m_1$ represents
the amplitude of a simple antiferromagnetic state, present inthe high
field part of the phase diagram.  Since the phase boundary between
high field phase and the paramagnetic phase has a positive slope
$({dB\over dT}>0)$ in the high field phase, $m_0$ must have an attractive
interaction with the mode $m_1$.  In general, $\twiddle m_0$, $\twiddle
m_1$ and $\twiddle m_2$ are all vectors.  We will omit this and also
any anisotropy energy terms; they are much smaller energy scales.  We
have for the free energy
$$F =-Jm_2^4 - J_1m_1^2 -J_2 m_0^2 m_1^2 - m_0B -k_B T\Sigma (S)\eqno
(3)$$

The various $m_i=m_0S_i$ where $m_0 = g\mu_\beta |S|$.  Thus the
amplitudes $S_0$, $S_1$ and $S_2$, at $T=0$ can be represented by
points on a unit circle in 3 dimensions.  In scaled variables
$$f = F/Jm_0^4 = -S_2^4 -g_1 S_1^2 -g_2 S_0^2 S_1^2 -bS_0 -t \Sigma
(S)\eqno(4)$$
Here $g_1=J_1/Jm_0^2$, $g_2=J_2/J$, $b=B/Jm_0^3$ and $t=k_bT/Jm_0^4$.
The amplitude $S$ is given by $S^2=S_0^2+S_1^2+S_2^2$.

At $b=t=0$, we can draw a phase diagram (see Fig.~1) in the plane
($g_1,g_2$) for the free energy in Eq.~(4).  At $t=0$, we have $S^2=1$.
For small $g_1$ and $g_2$, only $S_2$ can be non-zero.  Since $f$ is
monotonic in $S_2$, the minimum for $f$ must lie at $S_2=1$ so that
$f=-1$.  We call this region I.  In region II, $g_2$ is small and so we
take $S_0=S_2=0$.  Again since the minimum of $f$ (at $S_1=1$) must be
$f_{II}=-g_1$.  This minimum is lower than region I when $g_1>1$.  Thus
the phase boundary I--II must be $g_1=1$.

For larger values of $g_1$ and $g_2$, we expect $S_2=0$.  We can write
$$f_{III}= -g_1 (1-S_0^2) - g_2 S_0^2 (1-S_0^2)\eqno (5)$$
Its minimization with respect to $S_0$ leads to
$$S_0^2 ={1\over 2}\left( 1-{g_1\over g_2}\right ),\quad S_1^2 = 1-
S_0^2 ={1\over 2} \left (1+{g_1\over g_2}\right)\eqno(6)$$
and
$$f_{III}=-g_1 -{g_2\over 4} \left (1-{g_1\over g_2}\right )^2 =
-{(g_1+g_2)^2\over 4g_2}\eqno (7)$$

The boundary II--III, given by $f_{II}=f_{III}$ is therefore $g_1=g_2$.
The boundary I--III is given by
$$g_2= 2-g_1 +2\sqrt{(1-g_1)} \sim 4-2g_1+0(g_1^2)\eqno(8)$$

The point $(g_1,g_2)= (0,4)$ can also be derived from a stability
analysis.  The point $(g_1,g_2)=(1,1)$ is a confluence of three
different phases and may be called a triple point.  These results are
described in Fig.~(1).

At $t=0$, the problem is still solvable, to some extent.  For example,
we consider the case of the upper critical field.  In the ordered
state, $S_2=0$ and $S_1$ vanishes at the upper critical field.  The
transitions are all second order.  We have (on substituting
$S_0^2=1-S_1^2$, and expanding in powers of $S_1$)
$$f=-b-\left[ (g_1+g_2)-{b\over 2}\right] S_1^2 +
\left( g_2+{b\over 8}\right) S_1^4 + O(S_1^6)\eqno(9)$$

We see that an $S_1\neq 0$ solution exists only for
$b<2(g_1+g_2)=b_{c_2}$.  In order that the $b=0$ ground state be
$S_2\neq 0$, the $(g_1,g_2)$ parameters have to come from region I.\ \
For $g_2>1$, Eq.~(8) leads to an inequality,
$$b_{c_2}=2(g_1+g_2) \leq 4\sqrt{g_2}\eqno(10)$$

To calculate the lower critical field, $b_{c_1}$, we note that the field
causes the ground state to change from the one in region I to the one
in region II.  For a given set of $(g_1,g_2)$ the ground state in region II
contains a uniform magnetization $S_0$ given by Eq. ~(6).  In region II the
free energy is lowered by the Zeeman interaction  $-b\cdot S_0$.  To order b
the free energy in region I remains unchanged.  Equating the two free
energies, we get,
$$b_{c_1} = \sqrt{2g_2\over (g_2-g_1)} [1 - {(g_1+g_2)^2\over 4g_2}]\eqno(11)$$

The lower critical field is negligibly small near the zero field phase
boundary between regions I and III.  It also diverges at the line
$g_1 = g_2$.  Experimentally the value of the ratio $b_{c_2}/b_{c_1} \simeq 40$
and the parameters have to be rather close to the I-III phase boundary.
While $g_2 < 4$, $g_1$ must be quite small.

To determine the critical field when the transition to the high field
state becomes second order, we have to include the temperature $t\neq 0$.
Since the transition is second order, at least for $b > b_t$, a Ginzburg-
Landau expansion of the entropy is expected to be valid for $S_1$.
However the magnetization $S_0$ is large and the entropy expansion does
not work.

\vskip .15truein
\noindent{\sc III.  Phase Diagram}

We have chosen $(g_1,g_2)=(0.5,1.5)$ for illustration.  The ratio of
upper critical field to lower critical field for this choice is 7.\ \
This is far from the experimental (see e. g. Godfrin and Osheroff in
ref. ~3) value 47.  However, the qualitative
features are essentially independent of the exact value of $(g_1,g_2)$
in a given region in Fig.~(1).

It is not possible to determine $(g_1,g_2)$ from the $T=0$ critical fields
alone.  For a given ratio of $B_{c2}/B_{c1}$, we can determine a contour in
the $(g_1,g_2)$ plane.  If we take the experimental value of the $B_{c1} =
0.45T$, then $b_{c1}/t_c = 0.37$ using the exeprimental constants for the
nuclear
Bohr magneton.  Thus $b_{c1} = 0.53$, (for $t_c = 1.45$).  Using the ratio
for $B_{c2}/B_{c1} = 47$, we get $b_{c2} = 25$.  The model has an upper
limit of $b_{c2} < 8$ to guarantee that the ground state at $T = 0$
is the uudd state.  There is an easy resolution of this problem,
namely a g shift, all $b$'s are replaced by a $g_3 b$.  This however leaves us
unable to determine the parameters based on the phase diagram alone.
In any case there needs to be some accomodation of the finite temperature
renormalizations which in turn involves the mean field assumption on
the entropy.  We therefore are reluctant to use finite temperature
observables.  We hope to calculate temperature independent properties
such as spin wave velocities and the energies of defects to
fix the parameters at a later date.

At $b=0$, the transition occurs at $t_c=1.45$ into the uudd state.  The
entropy discontinuity is almost full in accord with ref.~(7).\ \  Note
that if the entropy $\Sigma(S)$ had been expanded in powers of $S$, as
noted in ref.~(8), the transition temperature would have been 2.93
$(t_c^{-1}= {1\over 12}+ {1\over \sqrt{15}})$ and the change in $S_2,
\Delta S_2=1.97$.\ \ In the present calculations $\Delta S_2=1$.  Since
$S_0$ and $S_1$ are zero in the ground state, the transition
temperature is independent of $(g_1,g_2)$.\ \ However the lower
critical field does depend on these parameters.

As the magnetic field increases, $t_c$ slowly decreases. There is a
point at which all three phases ($S_2$, $(S_0,S_1)$ and paramagnetic)
coexist, a triple point.  The transition between low field phase
($S_2$) and high field phase ($S_0,S_1$) is first order, as is the
transition between the paramagnetic and high field state.  These
experimental features appear to be reproduced by the model.  It came as
a surprise that the model also produces a critical point.  This is the
point $(b_t, t_t)$ at which the first order phase transition between disordered
paramagnetic phase $S_0 \neq 0, S_1 = 0$ and the high field phase
$(S_0, S_1) \neq 0$ becomes a second order phase transition.

We thus see in Fig. (2) a complete phase diagram for solid $^3$He
magnetism.  In region~I, the ground state is the well known
up-up-down-down state.  In region III, the ground state is $S_0\neq 0,
S_1\neq 0$.  Region II represents the disordered state.  Again the
choice of parameters determines the ratio $b_{c_2}/b_{c_1}\simeq 7$.
The transition at $b=0$ is independent of the parameters $(g_1,g_2)$
and is at 1.4.  The precise value of $b_{c_1}$ (at .6) clearly depends
on $(g_1,g_2)$.

In Fig. (2), the solid lines describe a first order phase transition
while the dashed line represents a second order transition.  Fig.~(3)
shows the discontinuity in $S_1(\Delta S_1)$ as a function of magnetic
field along the phase boundary.  We see that $\Delta S_1$ vanishes at
$b=1.25$.  For $b>1.25$, the phase transition is second order.  The
transition temperature increases with magnetic field until $t\simeq
2.2$ and $b=2.7$.  For $b>2.7$, the transition temperature decreases
with increasing magnetic field, reaching $t=0$ at
$b_{c_2}=2(g_1+g_2)=4$.
\vskip.15truein
\noindent{\sc IV. Summary}

We have shown that the phase diagram of solid $^3$He can be derived from
rather simple free energy considerations.  These are extensions of a
model proposed by Guyer and Kumar for the ordering transition at low
magnetic field.  The present results reproduce the essential features
of the phase diagram including the critical point in the high field-
paramagnetic phase boundary.

There are two possible directions for further work, both involving
the introduction of a space gradient dependent term in the free energy.
These are (1) the derivation of magnon (spin wave) dispersion and
(2) analysis of defects and their interaction with the magnons.
The dynamics of spins can be easily written down using the Bloch equations
for the motion of spins subjected to a local field derived from Eq. ~(4).
The defects are most likely domain walls between metastable states
frozen into the true ground state.  We will return to these questions later.

We acknowledge useful discussions with E. D. Adams, R. Guyer, M. Roger
and N. S. Sullivan.  This work was partially supported by the National
High Magnetic Field Laboratory and the US Department of Energy, grant
DEF G05-91-ER45462.

\vfill\eject

\noindent{\sc References}
\frenchspacing

\item{1.} There are several reviews of experiments on solid $^3$He with
references to original experiments.  See for example E.D. Adams, Can.
J. Phys. {\bf 65}, 1336 (1987) and D.D. Osheroff, J. Low Temp. Phys.
{\bf 87}, 297 (1992).  Also see D.E. Greywall and P. Busch, Phys. Rev.
{\bf B36}, 6853 (1987).

\item{2.} J.S. Xia, W. Ni and E.D. Adams, Phys. Rev. Lett. {\bf 70},
1481 (1993).

\item{3.} The multiple spin exchange model has been reviewed by M.
Roger, J. H. Hetherington and J.M. Delrieu, Rev. Mod. Phys. {\bf 55}, 1
(1983).  A more recent discussion can be found in D. D. Osheroff, H. Godfrin
R. Ruel, Phys. Rev. Lett. {\bf 58}, 2458 (1987) and H. Godfrin and
D.D. Osheroff, Phys. Rev. {\bf B38}, 4492 (1988).  An early discussion
can be found in D. J. Thouless, Proc. Phys. Soc. London {\bf 86}, 893
(1965); {\bf86}, 905 (1965).

\item{4.} See for example Osheroff (ref.(1)) and Greywall and Busch
(ref. (1)).

\item{5.} D. M. Ceperley and G. Jacucci, Phys. Rev. Lett. {\bf 58}, 1648
(1987), also see Godfrim and Osheroff op. cit. ref. (3).

\item{6.} Y.P. Feng, P. Schiffer, J. Mihalism and D. D. Osheroff, Phys.
Rev. Lett., {\bf 65}, 1450 (1990); Y.P. Feng, P. Schiffer and D. D.
Osheroff, Phys. Rev. B{\bf 49}, 8790 (1994).  These papers report
thermal conductivity due to scattering of magnons from umklapp
scattering and defects.  The properties of defects are one objective
in setting up the simple formalism reported here.

\item{7.} R.A. Guyer and P. Kumar, J. Low Temp. Phys. {\bf 47}, 321
(1982).

\item{8.} J. Dubois, B. Nelson and P. Kumar, UF preprint (1994).
\vfill\eject
\noindent{\sc Figure Captions}

\item{1.} The phase diagram in the $(g_1,g_2)$ parameter space at
$b=t=0$, zero magnetic field and temperature.  In region I, the only
non-zero order parameter is $S_2$, representing the up-up-down-down
state.  All other amplitudes $(S_0,S_1)$ are zero.  Similarly in region
II only $S_1\neq 0$, $S_0=S_2=0$.  In region III in contrast, $S_2=0$
and $S_0$ and $S_1$ are finite.

\item{2.} The phase diagram in the $(b,t)$ plane for $(g_1,g_2)=
(0.5,1.5)$.  The solid lines represent a first order phase transition
while the dashed line represents a second order transition.

\item{3.} The discontinuity in $S_1(\Delta S_1)$ at the phase
transition.  At small fields $(b\gtwid 0.5)$, the transition is clearly
hard second order since $\Delta S_1 \sim 0.6$.  However for $b>1.25$,
the transition is second order.

\bye